\newcommand{\as}{\mbox{$\alpha_{s}$}}
\newcommand{\mz}{\mbox{$m_{Z^0}$}}
\newcommand{\xp}{\mbox{$x_{p}$}}

\newcommand{\ycut}{\mbox{$y_{cut}$}}

\def\lms{\Lambda_{\overline{MS}}}
\def\mz{M_{Z^0}}
\documentstyle[12pt,psfig]{article}
\newlength{\dinwidth}
\newlength{\dinmargin}
\begin{document}
\renewcommand{\thefootnote}{\arabic{footnote}}
\newcommand{\sleq} {\raisebox{-.6ex}{${\textstyle\stackrel{<}{\sim}}$}}
\newcommand{\sgeq} {\raisebox{-.6ex}{${\textstyle\stackrel{>}{\sim}}$}}
%

\begin{titlepage}
\title{ Measurement of $\alpha_s$ from Jet Rates \\
 in Deep Inelastic Scattering at HERA }
\author{\rm ZEUS Collaboration}
\date{}
\maketitle
\vspace*{5.cm}

\begin{abstract}

\noindent
Jet production in deep inelastic scattering
for $120<Q^2<3600$~GeV$^2$ has been studied using data from an
integrated luminosity of 3.2~pb$^{-1}$ collected with the ZEUS
detector at HERA.  Jets are identified with the JADE algorithm.
A cut on the angular distribution
of parton emission in the $\gamma^*$-parton centre-of-mass system
minimises the experimental and theoretical uncertainties in the determination
of the jet rates.
The jet rates, when compared to ${\cal O}$(\as$^2$) perturbative QCD
calculations,
allow a precise determination of \as ($Q$) in three $Q^2$-intervals.
The values are consistent with a running of \as ($Q$), as expected from QCD.
Extrapolating to $Q=M_{Z^0}$ yields
$\as (\mz) = 0.117~\pm~0.005~(stat)~^{+0.004}_{-0.005}~(syst_{exp})~
{}~{\pm~0.007}~(syst_{theory})$.

\end{abstract}

\vspace{-19cm}
\begin{flushleft}
\tt DESY 95-182 \\
September 1995 \\
\end{flushleft}

\setcounter{page}{0}
\thispagestyle{empty}
\eject
\end{titlepage}


\def\3{\ss}
\footnotesize
\renewcommand{\thepage}{\Roman{page}}
\pagenumbering{Roman}
\begin{center}
\begin{large}
The ZEUS Collaboration
\end{large}
\end{center}
M.~Derrick, D.~Krakauer, S.~Magill, D.~Mikunas, B.~Musgrave,
J.~Repond, R.~Stanek, R.L.~Talaga, H.~Zhang \\
{\it Argonne National Laboratory, Argonne, IL, USA}~$^{p}$\\[6pt]
R.~Ayad$^1$, G.~Bari, M.~Basile,
L.~Bellagamba, D.~Boscherini, A.~Bruni, G.~Bruni, P.~Bruni, G.~Cara
Romeo, G.~Castellini$^{2}$, M.~Chiarini,
L.~Cifarelli$^{3}$, F.~Cindolo, A.~Contin, M.~Corradi,
I.~Gialas$^{4}$,
P.~Giusti, G.~Iacobucci, G.~Laurenti, G.~Levi, A.~Margotti,
T.~Massam, R.~Nania, C.~Nemoz, \\
F.~Palmonari, A.~Polini, G.~Sartorelli, R.~Timellini, Y.~Zamora
Garcia$^{1}$,
A.~Zichichi \\
{\it University and INFN Bologna, Bologna, Italy}~$^{f}$
% ...
\\[6pt]
\clearpage
\hspace*{1mm}
$^{ 1}$ supported by Worldlab, Lausanne, Switzerland \\
\hspace*{1mm}
$^{ 2}$ also at IROE Florence, Italy  \\
\hspace*{1mm}
$^{ 3}$ now at Univ. of Salerno and INFN Napoli, Italy  \\
\hspace*{1mm}
$^{ 4}$ supported by EU HCM contract ERB-CHRX-CT93-0376 \\
\hspace*{1mm}
% ...
$^{21}$ now with OPAL Collaboration, Faculty of Physics at Univ. of
        Freiburg \\
% ...

\begin{tabular}{lp{15cm}}
$^{a}$ & supported by the Natural Sciences and Engineering Research
         Council of Canada (NSERC) \\
$^{b}$ & supported by the FCAR of Qu\'ebec, Canada\\
$^{c}$ & supported by the German Federal Ministry for Education and
         Science, Research and Technology (BMBF), under contract
         numbers 056BN19I, 056FR19P, 056HH19I, 056HH29I, 056SI79I\\
$^{d}$ & supported by the MINERVA Gesellschaft f\"ur Forschung GmbH,
         and by the Israel Academy of Science \\
$^{e}$ & supported by the German Israeli Foundation, and
         by the Israel Academy of Science \\
$^{f}$ & supported by the Italian National Institute for Nuclear Physics
         (INFN) \\
$^{g}$ & supported by the Japanese Ministry of Education, Science and
         Culture (the Monbusho)
         and its grants for Scientific Research\\
$^{h}$ & supported by the Korean Ministry of Education and Korea Science
         and Engineering Foundation \\
$^{i}$ & supported by the Netherlands Foundation for Research on Matter
         (FOM)\\
$^{j}$ & supported by the Polish State Committee for Scientific
         Research, grants No.~115/E-343/SPUB/P03/109/95, 2P03B 244
         08p02, p03, p04 and p05, and the Foundation for Polish-German
         Collaboration (proj. No. 506/92) \\
$^{k}$ & supported by the Polish State Committee for Scientific
         Research (grant No. 2 P03B 083 08) \\
$^{l}$ & partially supported by the German Federal Ministry for
         Education and Science, Research and Technology (BMBF) \\
$^{m}$ & supported by the German Federal Ministry for Education and
         Science, Research and Technology (BMBF), and the Fund of
         Fundamental Research of Russian Ministry of Science and
         Education and by INTAS-Grant No. 93-63 \\
$^{n}$ & supported by the Spanish Ministry of Education and Science
         through funds provided by CICYT \\
$^{o}$ & supported by the Particle Physics and Astronomy Research
         Council \\
$^{p}$ & supported by the US Department of Energy \\
$^{q}$ & supported by the US National Science Foundation
\end{tabular}

%
%

\clearpage

\pagenumbering{arabic}
\setcounter{page}{1}
\normalsize

\section{Introduction}
\label{sec:intro}

Neutral current (NC) deep inelastic scattering (DIS)
($lp \rightarrow lX; l=e,\mu $), is characterised by the
exchange of a virtual photon or $Z^0$ boson between the incident
lepton and proton.
In the naive quark-parton-model (QPM) the process $V^* q \rightarrow q$ ($V =
\gamma,\,Z^0$) gives rise to 1+1 jets in the final state corresponding to
the struck quark from the proton and the proton remnant (hereafter denoted by
``+1'').  Multi-jet production in DIS beyond 1+1 jets provides a good
laboratory for testing quantum chromodynamics (QCD).
{}From the measured rate of 2+1 jet events it is possible to determine the
strong coupling constant \as , for fixed kinematics and a given jet
definition, by comparing to theoretical calculations in which \as\ is
the only free parameter.

To leading order in \as , 2+1 jet production proceeds via
QCD-Compton scattering ($V^* q \rightarrow qg$) and
boson-gluon fusion (BGF) ($V^* g \rightarrow q \overline{q}$).
For the extraction of \as\ from the measured jet rates to be reliable
the 2+1 jet rate must be calculated at least to next-to-leading order (NLO) in
QCD, where the renormalisation scheme is defined unambiguously.
Furthermore the jet definition has to be treated in the same way
in theory and experiment for a quantitative comparison with the
predictions of QCD.
Theoretical calculations \cite{Graudenz91,Brodkorb92,Graudenz94,Brodkorb94}
for the jet rates at the parton level are currently available only for the
JADE jet definition scheme \cite{JADE}.
Therefore the measured jet rates, obtained using the same jet-finding scheme,
have to be corrected to the parton level so that a comparison with
the NLO ${\cal O}$(\as$^2$) calculations can be made
in order to determine \as .
The extracted \as\ value can be expected to be reliable when
the NLO calculations reproduce the
corrected jet rates over a wide kinematic range and the extracted value is
insensitive to
the cuts applied at the detector level. In this analysis a cut on the
parton variable $z$ (described later) is applied, which restricts
the phase space so that these requirements are well satisfied.

Multi-jet production in DIS has been studied by the E665 fixed-target
experiment at FERMILAB at a low centre-of-mass
energy, $\sqrt s$, of $\sim$~30~GeV
\cite{E665}, and  at higher energies, $\sqrt s$=300~GeV, by
ZEUS \cite{ZEUS_cone} and H1 \cite{H1} at HERA where
jet structures are more clearly discernible.
This paper describes the extraction of $\as$ from measurements of multi-jet
rates at $Q^2$  between 120 and 3600~GeV$^2$.
An earlier study of jet rates and jet kinematics has
been reported by this experiment \cite{ZEUS_jet};
an extraction of \as\ from multi-jet production
has been reported by H1 \cite{H1_alphas}.

\section{The ZEUS Detector}
\label{sec:detector}

The data used in this analysis were collected with the ZEUS detector
during 1994 when HERA provided collisions between 27.5~GeV electrons or
positrons\footnote{Hereafter ``electron" is used in a generic sense to
refer to $e^-$ or $e^+$.} and
820~GeV protons, yielding a centre-of-mass energy of 300~GeV.
They correspond to an integrated luminosity of 3.2~pb$^{-1}$.

ZEUS is an almost hermetic, multipurpose, magnetic detector and has been
described elsewhere in detail \cite{detector}.
Here a brief description of the components relevant for this analysis is
given.
Charged particles are
tracked by the inner tracking detectors which operate in a
magnetic field of 1.43 T provided by a thin superconducting coil.
Immediately surrounding the beam pipe is the vertex detector,
a drift chamber, which
consists of 120 radial cells, each with 12 sense wires \cite{vxd}.
It is surrounded by
the central tracking detector which consists of 72 cylindrical
drift chamber layers,  organised into 9 `superlayers' \cite{ctd}.
In the present analysis these tracking detectors are primarily used for
the determination of the event vertex.

The energy associated with the hadronic final state and the scattered
electron is measured with the
uranium-scintillator calorimeter (CAL) \cite{CAL}
which consists of three parts: the
forward (FCAL), the rear (RCAL) and the barrel calorimeter (BCAL).The
ZEUS coordinate system is defined as right handed with the $Z$ axis pointing in
the proton beam direction, hereafter referred to as ``forward''. The   $X$ axis
points horizontally towards the centre of HERA and
the $Y$ axis points vertically upwards. The polar angle
$\theta$ is defined with respect to the  $Z$ direction.
Each part of the calorimeter is subdivided
longitudinally into one electromagnetic section (EMC) and one hadronic section
(HAC) for the RCAL and two HAC sections for BCAL and FCAL.   Holes of $20\times
20$ cm$^2$  at the centre of FCAL and RCAL accommodate the HERA beam pipe. In
the $XY$ plane around the FCAL beam pipe,  the HAC section is segmented in
$20\times 20$ cm$^2$ cells and the EMC section in $5\times 20$ cm$^2$ cells.
In total, the calorimeter consists of approximately 6000 cells. In terms of
pseudorapidity, $\eta=-\ln\tan \frac{\theta}{2}$, the FCAL covers the interval
$4.3\ge\eta\ge 1.1$, the BCAL  $1.1\ge\eta\ge -0.75$ and the RCAL
$-0.75\ge\eta\ge -3.8$, for the nominal interaction point at
$X=Y=Z=0$. The CAL energy resolution, as measured under test beam conditions,
is $\sigma_E/E=0.18/\sqrt{E}$ for electrons and $\sigma_E/E=0.35/\sqrt{E}$
for hadrons ($E$ in GeV). The time resolution of the
calorimeter, which is important for rejecting beam-gas backgrounds,
is better than 1 ns for energy deposits greater than 4.5 GeV.

\section{Event Kinematics}

For a given $ep$ centre-of-mass energy $\sqrt{s}$,
the differential cross section for
leading order ${\cal O}(\as^1)$   2+1 jet production
in DIS depends on 5 independent kinematic variables, which we take
as $x$, $y$, $x_{p}$, $z$, and $\Phi$ \cite{QCD_KMS_3JET_89}.
The first two, Bjorken-$x$ and $y$,
are sufficient to describe  the ${\cal O}(\as^0)$ QPM 1+1 jet process.
They correspond to the momentum fraction of the proton carried by the
struck quark ($x$) and the fractional energy transfer between the electron
and the proton in the proton rest frame ($y$).
Three additional variables ($x_p,z,\Phi$) are introduced to describe
the 2+1 parton kinematics.
The parton variable \xp\ is defined by:
$$   \xp=\frac {Q^2}{2~p\cdot q}= \frac{Q^2}{Q^2+m^2_{ij}}=\frac{x}{\xi},$$
where $q$ is the four-momentum
of the exchanged virtual boson in the $ep$ scattering process,
$\xi$ is the fraction of the proton's four-momentum $P$
carried by the incoming parton with four-momentum
$p = \xi P$, $m_{ij}$ is the invariant mass of the two non-remnant jets
and $Q^2=-q^2$. $Q^2,x$ and $y$ are related by $Q^2 = s~x~y$.
The parton variable $z$ is defined by:
$$ z_1 = \frac {p\cdot p_1}{p\cdot q} = \frac {1}{2} \cdot (1-\cos\theta^*_1)
 \  =  \ \frac{E_1 \cdot (1-\cos\theta_1)}{~\sum_{i=1,2} E_i \cdot
(1-\cos\theta_i)~}.$$
The formula is given for one of the partons $i=1$. The outgoing four-momentum
of the parton from the hard scattering is $p_1$ and $\theta^*_1$ is
the scattering angle in the
$\gamma^*$-parton  centre-of-mass system.
Experimentally, $z$ is determined in the HERA system
from the energies and angles, $E_i$ and $\theta_i$, of the two jets. The
jets are assumed to be massless.
The other parton satisfies the constraint $z_2=1-z_1$. The angle $\Phi$
represents the azimuthal angle between the parton and lepton scattering planes
in the  $\gamma^*$-parton centre-of-mass system.

Since the ZEUS detector is nearly hermetic, it is possible to reconstruct the
kinematic variables $x, y$
and $Q^2$ for NC DIS using different combinations of
the angles and energies of the scattered lepton and of the hadronic
system \cite{zeusf2}.
The electron
method was used to determine $y$ as $y_e$ from $E_e'$ and $\theta_e$,
the energy and polar angle of the
scattered electron. The hadronic, or Jacquet-Blondel method \cite{JB},
was used to reconstruct $y$ as $y_{JB}=\sum_{h}E_h(1-cos\theta_h)/(2E_e)$
where $E_h$ and $\theta_h$ are the energy and polar angle calculated
from the calorimeter cells not associated with the scattered electron,
and $E_e$ is the electron beam energy.
The double angle (DA) method uses $\theta_e$ and
$\gamma_H$, the polar angle of the struck quark in the QPM which is given by
$\cos\gamma_H = (\sum_{h}p^2_{T,h} - (2 E_e y_{JB})^2)/
                  (\sum_{h}p^2_{T,h} + (2 E_e y_{JB})^2)$.
The DA method, which  measures $Q^2$ with small bias and good resolution in
the kinematic range of this analysis, was
used to reconstruct the $x,y$ and $Q^2$ variables of NC events
(and the jet variables $x_p$ and $z$ defined above) \cite{zeusf2}.

\section{Trigger Conditions and Event Selection}
\label{sec:selection}

The trigger and event selection followed closely
that described in reference \cite{ZEUS_jet}.
The trigger acceptance was essentially independent of the hadronic final
state with an acceptance greater than 97~\% for $Q^2>10$ GeV$^2$.
Neutral current DIS events were selected by the following criteria:
the event times measured by the FCAL and the RCAL had to be consistent
with an interaction inside the detector. This cut strongly reduced beam-gas
background. The $Z$ position of the event vertex was reconstructed from the
tracking data. Events were accepted if the $Z$ position was within
$\pm$~50~cm of the nominal interaction point.
An electron candidate
with energy greater than 10 GeV had to be found in the calorimeter.
To reject backgrounds from
photoproduction events with a fake electron (mostly $\pi^0$'s
close to the proton beam) the electron candidate was required
to satisfy $y_e < 0.95$.
Photoproduction and beam-gas backgrounds were further
suppressed by demanding energy-momentum conservation.
For a fully contained event, and neglecting the detector resolution,
one expects $E-P_Z = 2\cdot E_e$,
where $E$ and $P_Z$ are the summed energy and $Z$-component of the
momenta of all objects 
measured in the calorimeter.
Taking the detector resolution into account 35~GeV$<E-P_Z<$~60~GeV was
required to select DIS events.
The background from photoproduction was estimated to be negligible
and that from QED Compton scattering was found to be less than 1\%.
Diffractive events, which do not deposit a significant amount of energy in
the FCAL, did not pass the selection criteria given below.
Finally, beam halo muons and cosmic rays were rejected by
suitable algorithms.

Several considerations motivated the selection of the kinematic region used
for the \as\ measurement.    First,
the analysis was restricted to high $Q^2$, where clear jet structures are
observed and hadronisation uncertainties are minimised.
Secondly, theoretical uncertainties in the 2+1 jet cross section are
small at high $x$,
where the parton densities of the proton are well known.
In addition at high $x$ the uncertainty stemming from the initial state
parton-showers,
used in the Monte Carlo simulation to correct the data to
the partonic level, was reduced.
Thirdly, the acceptance for 2+1 jet events increases at high $y$: in
particular, the
forward jet is well contained within the detector.
Finally, all of the above concerns were balanced
against the need for sufficient statistics.
The kinematic region selected
for the final analysis was therefore:  $120<Q^2<3600$~GeV$^2$, $0.01<x<0.1$,
and $0.1<y<0.95$, resulting in a sample of 4472 events. The $Q^2$
range was further subdivided into three regions to measure \as $(Q)$ at
increasing scales as a consistency check and as a test for the running of
the strong coupling constant.  These ranges were:
$120<Q^2<240$, $240<Q^2<720$, and $720<Q^2<3600$~GeV$^2$.
The number of events in each region were
1649, 2048 and 775, respectively.

\section{Jet Definition and Jet Kinematics}

The JADE algorithm \cite{JADE} was used to relate the hadronic final state
measured in the detector to the underlying hard scattering processes.
It is a cluster algorithm based on the
scaled invariant mass,
$y_{ij} = m_{ij}^2/W^2 = 2 E_i E_j (1-cos\theta_{ij}) / W^2$,
where $m_{ij}$ is the invariant
mass of the two objects $i$ and $j$, which are assumed to
be massless. The scale $W^2$
is the squared invariant mass of the overall hadronic final state and
$E_i, E_j$ and $\theta_{ij}$ are the energies of the objects and the angle
between them.
Starting with the minimum $y_{ij}$ of all possible combinations,
objects were merged  by adding
their four-momenta until $y_{ij}$ for all objects exceeded a jet resolution
parameter \ycut. Those objects remaining were then considered as jets.
The JADE algorithm was slightly modified \cite{MOD_JADE,MC_ANWENDUNG}
for use at the detector level in $ep$ collisions by the addition of a
pseudo-particle inserted along the $Z$ axis.
The missing longitudinal momentum in each event was assigned to the
momentum of the pseudo-particle. It prevents the detected fraction
of particles originating from the proton remnant
from forming spurious jets.

The measured calorimeter energies
above 150 (200) MeV for EMC (HAC) cells and their angles relative
to the interaction point were used to define vectors which were input to the
JADE algorithm in the detector level analysis.
At the detector level, the scale $W$ was calculated as
$W^2_{vis}=s\,(1-x_{DA})\,y_{JB}$.
This value reflects the measured rather than
the true hadronic activity
and so reduces the event-by-event correction for the  detector resolution
when calculating $y_{ij}$.

\begin{figure}[p]
\vspace{-9pt}
\centerline{ \hspace{0.5cm}
             \psfig{figure=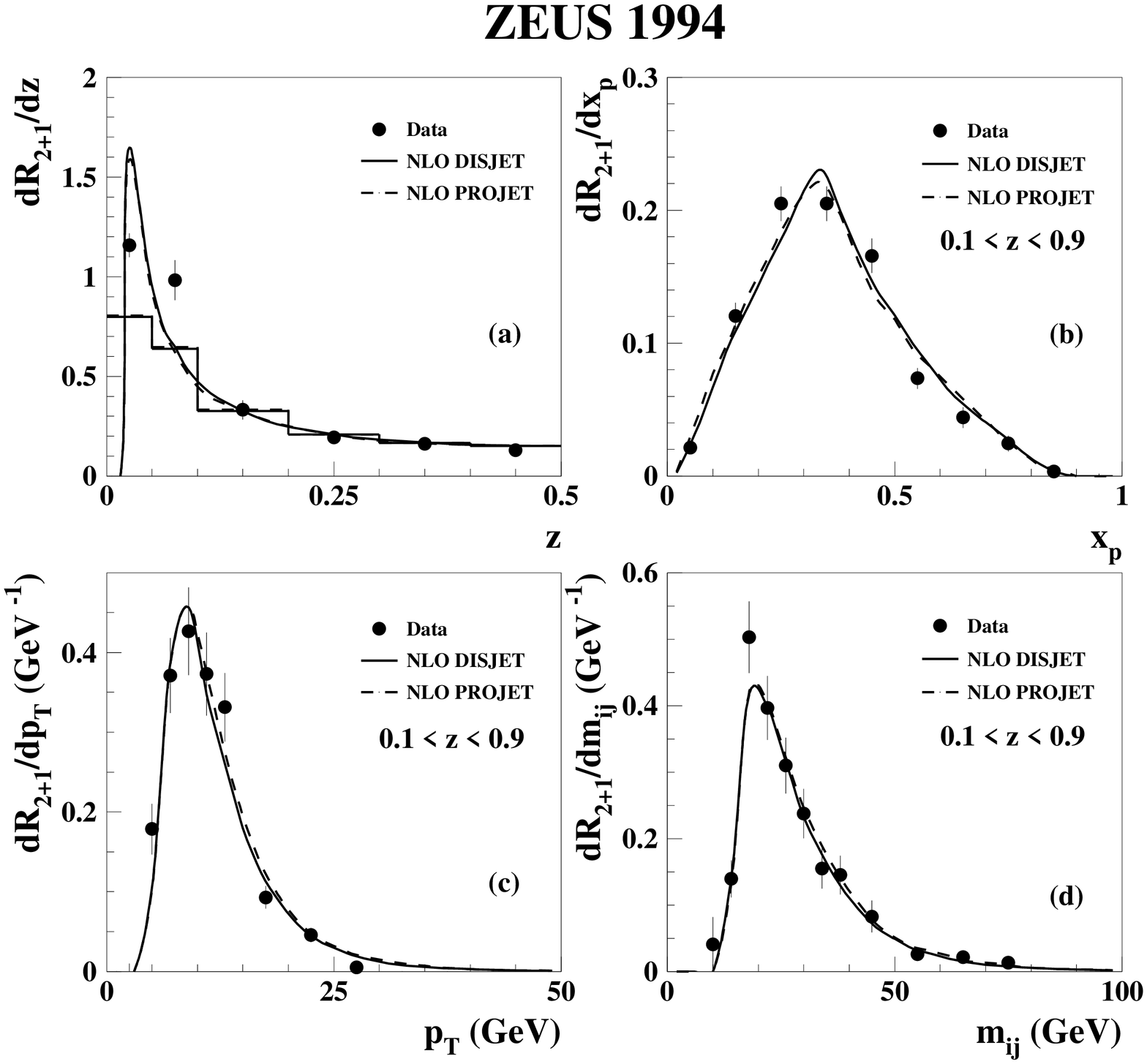,height=7.5in,width=7.0in}}
\vspace{-1.0cm}
\caption{ (a) Distribution of the parton variable, $z$,
of one of the two non-remnant jets in 2+1 jet events
in the range $120<Q^2<3600$ GeV$^2$, compared to the NLO calculations
(PROJET and DISJET).
The dots with error bars are the measured data.
The curves represent
the theoretical predictions after the application of the cuts.
The histograms show the same theoretical prediction with the binning
of the data.
(b) Distribution of $x_p$ for 2+1 jet events.
(c) Transverse momentum distribution $p_T$ for the two jets.
(d) Invariant mass distribution $m_{ij}$ of the two non-remnant jets.
Only events satisfying $0.1<z<0.9$ were plotted in Figs.~b--d.
All jet rates are evaluated for \ycut~=~0.02.
The data points are corrected to the parton level
and plotted with their statistical errors only. }
\label{fig:zxp}
\end{figure}

For 2+1 jet production in DIS, one of the non-remnant jets is typically
directed forward  because of the forward singularity in the cross section.
Such forward singularities are regulated in the theory by the cutoff
\ycut. Requiring a large \ycut\ is, however, not sufficient to avoid the
problems arising from forward-going jets close to the
beam pipe and the proton remnant.
This is achieved by a cut on the
parton variable, $z$, for 2+1 jet events\footnote{$z$ is
not defined for 1+1 and 3+1 jet events. In the following, 2+1 jet events
failing to pass the $z$ cut are not considered as 2+1 jet events. 1+1 and
3+1 jet events are counted in $N_{tot}$, the total number of selected
events.}. In QCD the rapid
rise towards $z=0$ results from collinear and infrared
singularities. In order to avoid this kinematical region the analysis
was restricted to events satisfying $0.1<z<0.9$.
This requirement also reduces the fraction
of forward jets ($\theta_{jet}<8^\circ$) from 30\% to 10\%.
Figure~\ref{fig:zxp}a shows the $dR_{2+1}/dz$ distribution for 2+1 jet events.
Here $R_{j+1}=N_{j+1}/N_{tot}$,
where $j$ stands for 0, 1, 2, or 3, $N_{j+1}$ is the number of (j+1)
jet events and $N_{tot}$ is the
total number of selected DIS events.
Figures~\ref{fig:zxp}b--d show the resulting
$x_p$, $p_T$ and $m_{ij}$ distributions.
The predictions of the NLO calculations of DISJET and PROJET (discussed
later) are also shown in Fig.~\ref{fig:zxp}.
The $z$-cut results in a significantly improved
agreement between the calculations and the data compared to our
earlier analysis done without this restriction on $z$ \cite{ZEUS_jet}.
This cut removes jets with transverse momenta $p_T$ below $\sim$~4 GeV
where $p_T$ is measured with respect to the
$\gamma^*$ direction and is calculated in the $\gamma^*$-parton system as:
$$ p_T = \sqrt {~ Q^2 \cdot \frac{1}{x_p} \cdot (1-x_p) \cdot z~(1-z) }.$$

\section{Jet Reconstruction and Jet Rates}

The acceptance and resolution,
as well as the correction of the measured jet rates to the parton level,
were determined using Monte Carlo methods.
Neutral current DIS events, generated
using LEPTO~6.1 \cite{LEPTO} and the Lund string fragmentation
model \cite{Lund} for the hadronisation, were
interfaced via DJANGO6 2.1 \cite{DJANGO} to
HERACLES~4.1 \cite{HERACLES} for QED radiative corrections. They were passed
through a GEANT \cite{geant} based detector
simulation, and subsequently analysed with the same reconstruction, selection
and jet analysis procedures as the data.  Both the hard
emission of partons at the matrix element level (calculated to leading order
in \as ) and the higher order soft parton showers are included in
the LEPTO matrix element, parton shower (MEPS) model.
The MEPS model satisfactorily describes the global jet
properties and production rates observed for the data
in the selected kinematic region \cite{ZEUS_jet}.

When generating events with the MEPS model, default values of all parameters
were used except for the
parameter $y_{min}$, which sets a minimum $y_{ij}$ of partons in first
order QCD matrix elements \cite{LEPTO}.
The value of $y_{min}$  was lowered  from
0.015 to 0.005 in order to study the measured jet rate as a function of
the jet resolution parameter \ycut\ for \ycut~$>$0.01.
The parton densities of the proton were taken from the
MRSD$'$- set \cite{MRSD}.

The jets were reconstructed by applying the JADE algorithm
at the parton level, the hadron level and the detector level.
These jets were constructed respectively from the output of the parton
shower step of the
event generator, the true momenta of the hadrons before the detector
simulation and the energy deposits in the calorimeter cells after the
detector simulation. The ratio of the jet rates at the
different levels of the event simulation are the corresponding correction
factors for hadronisation ($C_h$) and detector simulation ($C_d$), with
which the measured jet rates were multiplied.
Both the detector and hadronisation corrections
were found to be below 20\%.
Table~\ref{RJ_TABLE} shows the correction factors and
the corrected jet rates
$R_{2+1}$ for the three $Q^2$ intervals and for the combined
$Q^2$ region.

{\footnotesize
\begin{table}[htb]
\begin{center}
\begin{tabular}{|c||r|r|r|r|r|r|r|r|r||r|r|r|}
\hline
                  & \multicolumn{3}{c|}{$120<Q^2<240$}    &
                    \multicolumn{3}{c|}{$240<Q^2<720$}   &
                    \multicolumn{3}{c||}{$720<Q^2<3600$} &
                    \multicolumn{3}{c|}{$120<Q^2<3600$}   \\
                  & \multicolumn{3}{c|}{ (~GeV~$^2$) }             &
                    \multicolumn{3}{c|}{ (~GeV~$^2$) }             &
                    \multicolumn{3}{c||}{ (~GeV~$^2$) }            &
                    \multicolumn{3}{c|}{ (~GeV~$^2$) }              \\
\hline
 $y_{cut}$ & $R_{2+1}$~~ &$C_d$ &$C_h$ &$R_{2+1}$~~ &$C_d$ &$C_h$
     &$R_{2+1}$~~ &$C_d$ &$C_h$  &$R_{2+1}$~~ &$C_d$ &$C_h$  \\ \hline
 0.010  & 12.1$\pm$0.9 &1.02 &1.04 & 13.5$\pm$0.9 &0.92 &1.04
                & 11.5$\pm$1.3 &0.88 &1.03 & 12.6$\pm$0.6 &0.94 &1.05 \\
 0.015  & 10.0$\pm$0.8 &0.99 &1.04 & 10.8$\pm$0.8 &0.96 &1.05
                &  9.3$\pm$1.2 &0.89 &1.02 & 10.4$\pm$0.5 &0.96 &1.05 \\
 0.020  &  7.8$\pm$0.7 &0.96 &1.05 &  9.0$\pm$0.7 &0.96 &1.05
                &  8.6$\pm$1.1 &0.92 &1.01 &  8.6$\pm$0.5 &0.96 &1.05 \\
 0.030  &  5.3$\pm$0.6 &0.92 &1.10 &  6.5$\pm$0.5 &0.98 &1.08
                &  6.7$\pm$1.0 &0.95 &1.05 &  6.2$\pm$0.5 &0.96 &1.08 \\
 0.040  &  4.1$\pm$0.6 &0.93 &1.13 &  4.6$\pm$0.5 &0.96 &1.10
                &  4.6$\pm$0.8 &0.94 &1.06 &  4.4$\pm$0.4 &0.96 &1.08 \\
 0.050  &  3.3$\pm$0.5 &1.02 &1.16 &  3.6$\pm$0.5 &0.99 &1.11
                &  3.9$\pm$0.8 &0.93 &1.07 &  3.5$\pm$0.3 &0.99 &1.10 \\
 0.060  &  2.3$\pm$0.4 &0.94 &1.20 &  2.7$\pm$0.3 &1.02 &1.15
                &  2.9$\pm$0.7 &0.92 &1.07 &  2.6$\pm$0.2 &0.99 &1.13 \\
\hline
\end{tabular}
\end{center}
  \caption{ 2+1 jet production rates (in \%) corrected to the parton
level ($R_{2+1}$) and correction factors for
detector effects ($C_d$) and for hadronisation ($C_h$) in the
three $Q^2$ intervals and for the combined region. Errors
shown are statistical only. }
  \label{RJ_TABLE}
\end{table}
}

\section{${\cal O}(\as^2)$ Perturbative QCD Calculations }
\label{sec:qcd}

For NC electron-proton scattering, the 2+1 jet differential cross section
at the ${\cal O}(\as^1)$ LO level, expressed in terms of the above variables,
is given by:

$$ { {d^2 \sigma_{2+1}} \over {dx~dy}} =
    {{\alpha^2~\as} \over {y~Q^2}} \cdot \int {{dx_p} \over {x_p}} ~
               \int {dz} ~ \int {{d\Phi} \over {2\pi}} ~ (I_g + I_q),$$
where $I_g$ and $I_q$ are the gluon- and quark-initiated contributions
respectively \cite{QCD_KMS_3JET_89}
which contain singularities at
$z=0, \ z=1$ and $x_p=1$. The singularities correspond to the
limit where two partons are irresolvable due to
vanishing energy of one of the partons or vanishing opening angle.
In the BGF process the singularity is related to the collinear emission
of an outgoing quark in the proton remnant direction;
for the QCD Compton process the singularity occurs
when the momentum of the gluon is parallel to that of the quark
or when the quark emits a very soft gluon.
The integration must be done
numerically because of the $x$-dependent parton density distributions.
The integration limits for $z$ and $x_p$ in the JADE scheme
are functions of the scaled invariant mass cutoff $y_{cut} = m_{ij}^2/W^2$,
where $W$ is the reference mass scale and $m_{ij}$ is the invariant
mass between any two partons \cite{Brodkorb92}.
Any pair of partons with a scaled invariant mass below this
cutoff is not resolved. Therefore the singularities are regulated
by a single cutoff parameter $y_{cut}$. The leading order (LO) and
next-to-leading (NLO) order  2+1 jet cross section
can be expressed as:
$$ \frac {d\sigma_{2+1}^{LO}} {dx~dy} = c_{31} \cdot \alpha_s,$$
$$ \frac {d\sigma_{2+1}^{NLO}}{dx~dy} = c_{31} \cdot
                   \alpha_s + c_{32}\cdot \alpha_s^2.$$

Next-to-leading order corrections to $d\sigma_{2+1}$ include the
contribution from unresolved 3+1 jet  events as well as
negative corrections coming from virtual
loops \cite{Graudenz94,Brodkorb94}.
The coefficients $c_{ij}$ contain the hard scattering matrix elements and
the parton density functions of the incoming proton. The effect of a
change in \as\ on the parton densities is negligible
for our $Q^2$ range \cite{VOGT}.
The first index, $i$, stands for the jet multiplicity (including the
remnant jet) and the second index, $j$, represents the order of the
\as\ calculation.
After integrating over the jet variables ($x_p,z,\Phi$)
the coefficients $c_{ij}$ are functions of the event kinematic variables
$x,y$, \ycut \ and the factorisation scale $\mu_F$;
$c_{32}$ depends also on the renormalisation scale $\mu_R$ \cite{ingelman_rs}.
The parton densities contained in $c_{ij}$ are calculated at the scale $Q^2$.
In finite order perturbative QCD calculations \as\ depends on the
renormalisation scale $\mu_R$.
The 2+1 jet rates are derived from the cross sections by
$R_{2+1} = \sigma_{2+1} / \sigma_{tot}$.
The resulting ${\cal O}(\as^2)$ corrections to $R_{2+1}$,
using the NLO calculations, are considerable:
they vary from --20 to $+20$\%  when \ycut\ is varied between 0.01 and 0.06
in the kinematic region used for this study \cite{Brodkorb94,Graudenz94}.
The numerical cross section calculations are available in the DISJET
program \cite{DISJET} by Brodkorb and Mirkes, and in the PROJET
program \cite{PROJET} by Graudenz.
Both programs agree in their predictions of \as\ for a given
jet rate and they reproduce the shape of the measured jet rate
distributions as a function of \ycut \ well in the investigated kinematic
range (see below).

The renormalisation scheme used in the calculation is the $\overline{MS}$
scheme.
In second order, the dependence on other renormalisation schemes can
be completely specified by one parameter, which can be chosen to be the
value of the renormalisation scale $\mu_R$.
We chose $\mu^2_R=Q^2$ for our analysis.
The same scale is chosen for the factorisation scale $\mu_F$.
The parton densities were calculated with a fixed $\lms^{(5)} = 154$~MeV.
In the kinematic
range used in this analysis, the effect of varying $\lms$\ in the parton
densities is expected to be small \cite{VOGT}.
In the programs the contributions from the $c-$ and $b-$quarks
are zero in the parton density parametrisations below the single
quark mass thresholds, as defined in the $\overline{MS}$ factorisation
scheme. Above threshold, the contributions from the $c-$ and $b-$quarks
are calculated assuming zero mass.
The number of flavours in the formula for the running coupling constant is
changed at the single mass threshold, as required by the
$\overline {MS}$ renormalisation scheme, giving rise to five flavours
for $\lms$\ if $Q^2> m^2_b$.
At the BGF vertex we have used five flavours too because in the kinematic
range of this analysis $m^2_{ij}$ is above $4 \cdot m^2_b$.
Using four flavours at the BGF vertex in the PROJET program
would increase the $\as(\mz)$ value by 0.0025.
In the $x,Q^2$ region under study the contribution from massive $b-$quarks
to the proton structure function is calculated to be below 2\%  \cite{GRS}.

\section{Determination and $Q^2$ Dependence of \as}
\label{sec:jet_analysis}

The value of \as\ was determined by varying $\lms^{(5)}$ in the QCD
calculation until the best fit to the ratio $R_{2+1}$ was obtained
at \ycut~=~0.02.
The slope of the measured $R_{2+1}$ as a function of \ycut\
agrees with that from the calculation, showing that the result is not
sensitive to the particular value of \ycut\ used.
We chose \ycut~=~0.02 for the fit because the contribution from
$R_{3+1}$, which is a higher order effect,
becomes negligible for \ycut~$\geq$~0.02.
Furthermore the statistics are large and the jets are resolvable at
this value and the 2-jet system has a large invariant mass.

\begin{figure}[p]
\vspace{9pt}
\centerline{ \hspace{0.5cm}
             \psfig{figure=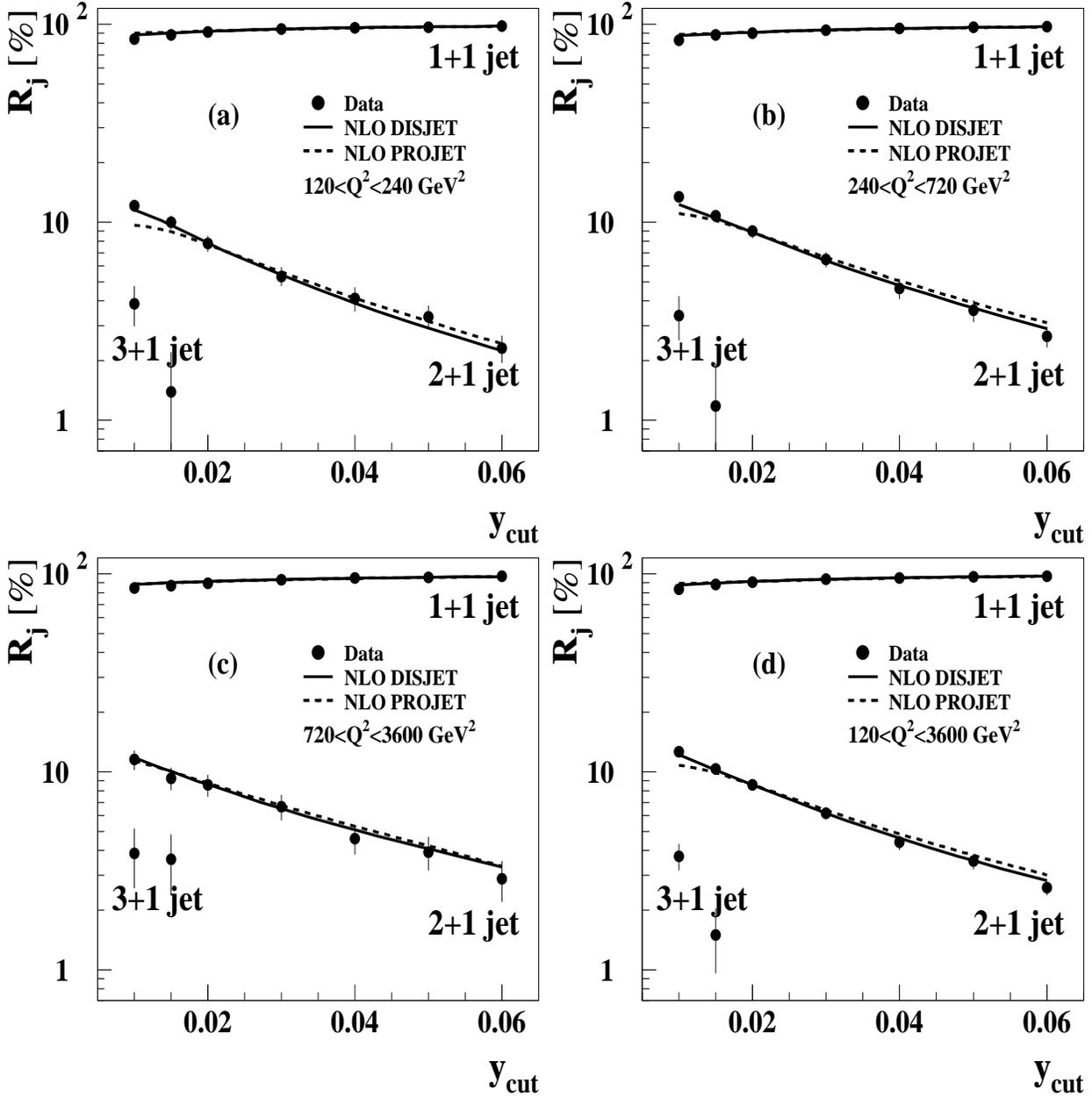,height=8.0in,width=7.0in}}
\caption{ Jet production rates $R_j$ as a function of the jet
          resolution parameter \ycut\ for $Q^2$ in the range
          (a) $120<Q^2<240$~GeV$^2$, (b) $240<Q^2<720$~GeV$^2$,
          (c) $720<Q^2<3600$~GeV$^2$, and (d) $120<Q^2<3600$~GeV$^2$.
          Only statistical errors are shown.
          Two NLO QCD calculations, DISJET and PROJET,
          each with the value of $\lms$ obtained from the fit at
          \ycut=0.02, are also shown.
        }
\label{fig:jet_rates}
\end{figure}

Figures~\ref{fig:jet_rates}a--d show the corrected jet rates,
$R_{1+1}, ~ R_{2+1}$ (also shown in Tab.~\ref{RJ_TABLE})
  and $R_{3+1}$ as a function of
\ycut\ for data compared with the DISJET and PROJET NLO QCD calculations
for the three $Q^2$ intervals and for the combined region.
Only statistical errors are shown.
All NLO terms are taken into account in both programs;
however, they use different approximations for some of these terms.
There is good agreement between the
corrected jet rate and the NLO QCD calculation over most of the
range in \ycut\ shown and in particular at the nominal \ycut~=~0.02,
where the \as\ value was extracted for this analysis.
Both programs agree well in their prediction of the jet-rate dependence
as a function of \ycut.
The best fit values for \as\
are used in the calculation. The range in \ycut\
was restricted to 0.01 to 0.06 because at lower values of
\ycut\ the jets are not experimentally resolvable  and
higher order corrections are significant,
while at larger values terms proportional to \ycut,
neglected in the calculation, become significant.
Moreover, uncertainties in the renormalisation scale and the
hadronisation corrections also become
large for \ycut\ above $\approx 0.06$ \cite{ingelman_rs}.

\begin{figure}[p]
\vspace{9pt}
\vspace{-1.8cm}
\centerline{ \psfig{figure=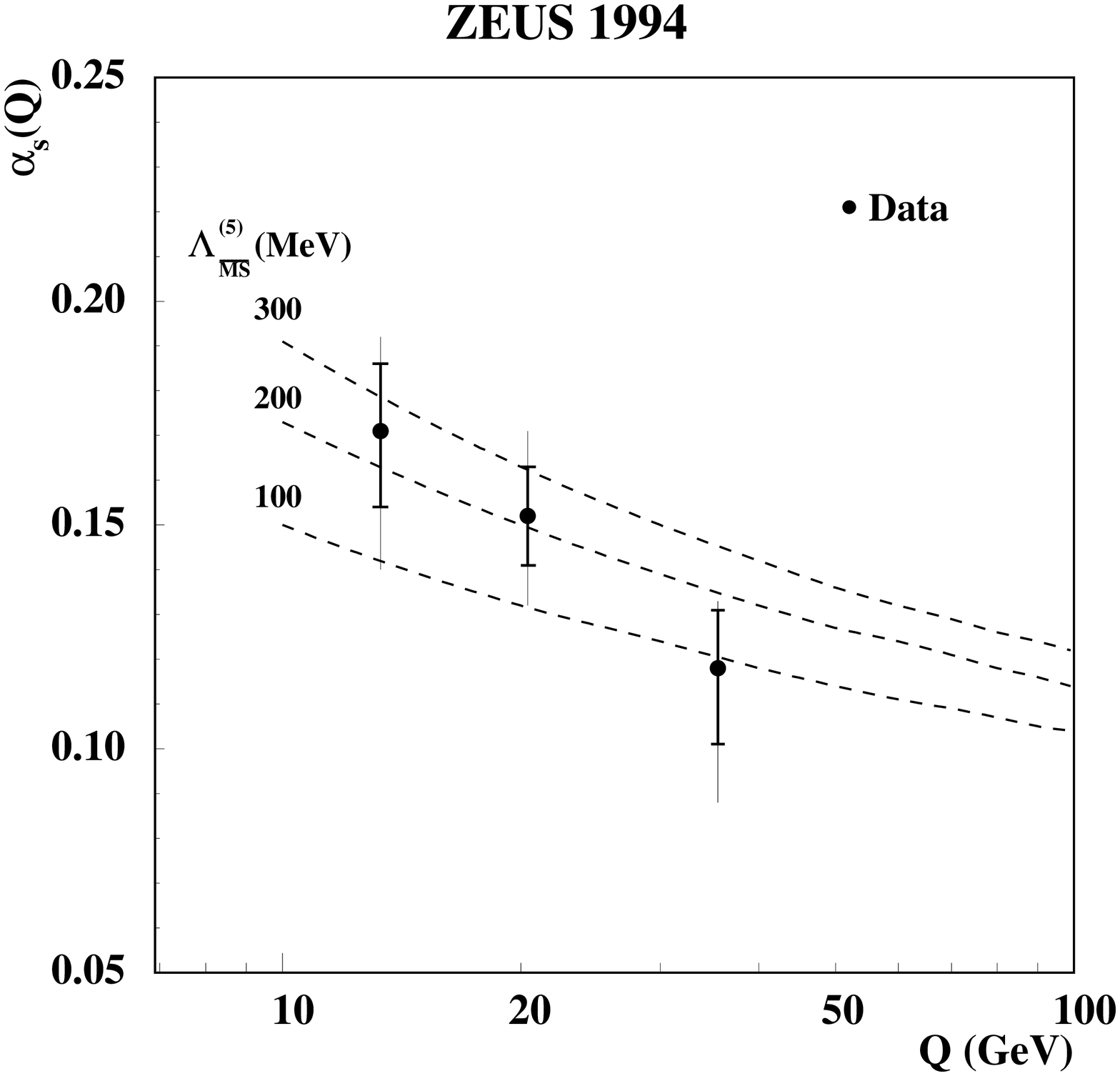,height=6.5in,width=6.5in}}
\caption{ Measured values of $\as(Q)$ for three different $Q^2$ regions.
The statistical error corresponds to the inner bar and the thin bar reflects
the statistical and systematic error added in quadrature.
Note that the systematic errors are strongly correlated.
The dashed curves represent \as\ with
$\lms^{(5)}= 100$, 200, and 300~MeV.
}
\label{fig:as_q2}
\end{figure}

The values of \as\ are plotted in Fig.~\ref{fig:as_q2}
as a function of $Q$ for each of the three $Q^2$ ranges.
They are calculated from the fitted values of $\lms^{(5)}$.
In Tab.~\ref{tab:as} the \as\ values determined for the three ranges in
$Q^2$ as well as for the full kinematic range are listed. Also shown are
the values of \as\ extrapolated to $Q=\mz$.
Both statistical and systematic
errors (discussed in the following section) are given.

In addition Fig.~\ref{fig:as_q2} shows the curves for
$\lms^{(5)}$ = 100, 200, and 300~MeV.
The measured \as\ decreases with increasing $Q$, consistent
with the running of the strong coupling constant if $Q^2$ is taken as the
scale. The fit to
a running \as\ (where \as\ was determined in the full $Q^2$ range)
yields a $\chi^2$ of 2.2 for 2 degrees of freedom, which
corresponds to a confidence level of 58.6\%.
A least squares fit to the hypothesis of a constant \as\ was performed.
Only statistical errors were considered in this fit
as the systematic errors are strongly correlated.
This fit yields
a $\chi^2$ of 7.7 for 2 degrees of freedom, which corresponds to
a confidence level of 2.1\%.
Taking into consideration the systematic uncertainties,
the $\chi^2$ for constant \as\ varies from 4.4 (changing energy scale
by -5\%) to 10.3.
A constant \as\ is thus ruled out at 90\% confidence level.
The three values of \as , expressed at the mass of the $Z^0$ boson, are
consistent within the errors.

\begin{table}[htb]
\centerline{
\begin{tabular}{|c|c|c|c|c|}
\hline
  $Q^2$ & $<Q>$ & $\lms^{(5)}$ & \vspace*{0.2cm} $\as(Q)$ \vspace*{-0.2cm}
                               & \vspace*{0.2cm} $\as(\mz)$ \vspace*{-0.2cm} \\
        (~GeV$^2$~) & (~GeV~)  &   (~MeV~)      &          &                 \\
\hline
     & & & & \\
120 --~~240
     & 13.3
     & $251~  ^{+108}_{\ -97}   $ $^{+31}_{-74}  $ $^{+115}_{-105}$
     & $0.171~ ^{+0.015}_{-0.017}$ $^{+0.005}_{-0.012}$ $^{+0.016}_{-0.018}$
     & $0.120~ ^{+0.007}_{-0.008}$ $^{+0.002}_{-0.006}$ $^{+0.007}_{-0.009}$\\
     & & & & \vspace{-0.1cm} \\
240 --~~720
    & 20.4
    & $217~  ^{+90}_{-74}    $ $^{+76}_{-60}    $ $^{+119}_{\ -67}$
    & $0.152~ ^{+0.011}_{-0.011}$ $^{+0.010}_{-0.009}$ $^{+0.014}_{-0.010}$
    & $0.117~ ^{+0.006}_{-0.007}$ $^{+0.006}_{-0.005}$ $^{+0.008}_{-0.006}$ \\
    & & & & \vspace{-0.1cm} \\
720 -- 3600
    & 35.5
    & $86~   ^{+82}_{-58}    $ $^{+30}_{-47}    $ $^{+61}_{-24}$
    & $0.118~ ^{+0.013}_{-0.017}$ $^{+0.006}_{-0.012}$ $^{+0.010}_{-0.006}$
    & $0.103~ ^{+0.010}_{-0.013}$ $^{+0.004}_{-0.010}$ $^{+0.008}_{-0.004}$\\
    & & & & \vspace{-0.1cm} \\
\hline
    & & & & \vspace{-0.1cm} \\
120 -- 3600
    &  22.1
    &  $208~  ^{+64}_{-53}    $ $^{+57}_{-50}   $ $^{+89}_{-75}$
    &  $0.148~ ^{+0.008}_{-0.008}$ $^{+0.007}_{-0.007}$ $^{+0.011}_{-0.012}$
    &  $0.117~ ^{+0.005}_{-0.005}$ $^{+0.004}_{-0.005}$ $^{+0.007}_{-0.007}$\\
    & & & &  \vspace{-0.01cm} \\
\hline
\end{tabular}
        }
\caption{
The measured values of $\lms^{(5)}$ and \as\ for the three ranges in $Q^2$
as well as for the full $Q^2$ range.
The first error is statistical, the second corresponds to the
experimental systematic uncertainty and the third to the theoretical
systematic uncertainties (hadronisation, parton density and scale
uncertainty).
        }
\label{tab:as}
\end{table}

\section{Systematic Uncertainties}
\label{sec:systematics}

\subsection{Experimental, Hadronisation and Parton Density Effects}

\begin{figure}[p]
\vspace{-1.0cm}
\vspace{-0.8cm}
\centerline{ \psfig{figure=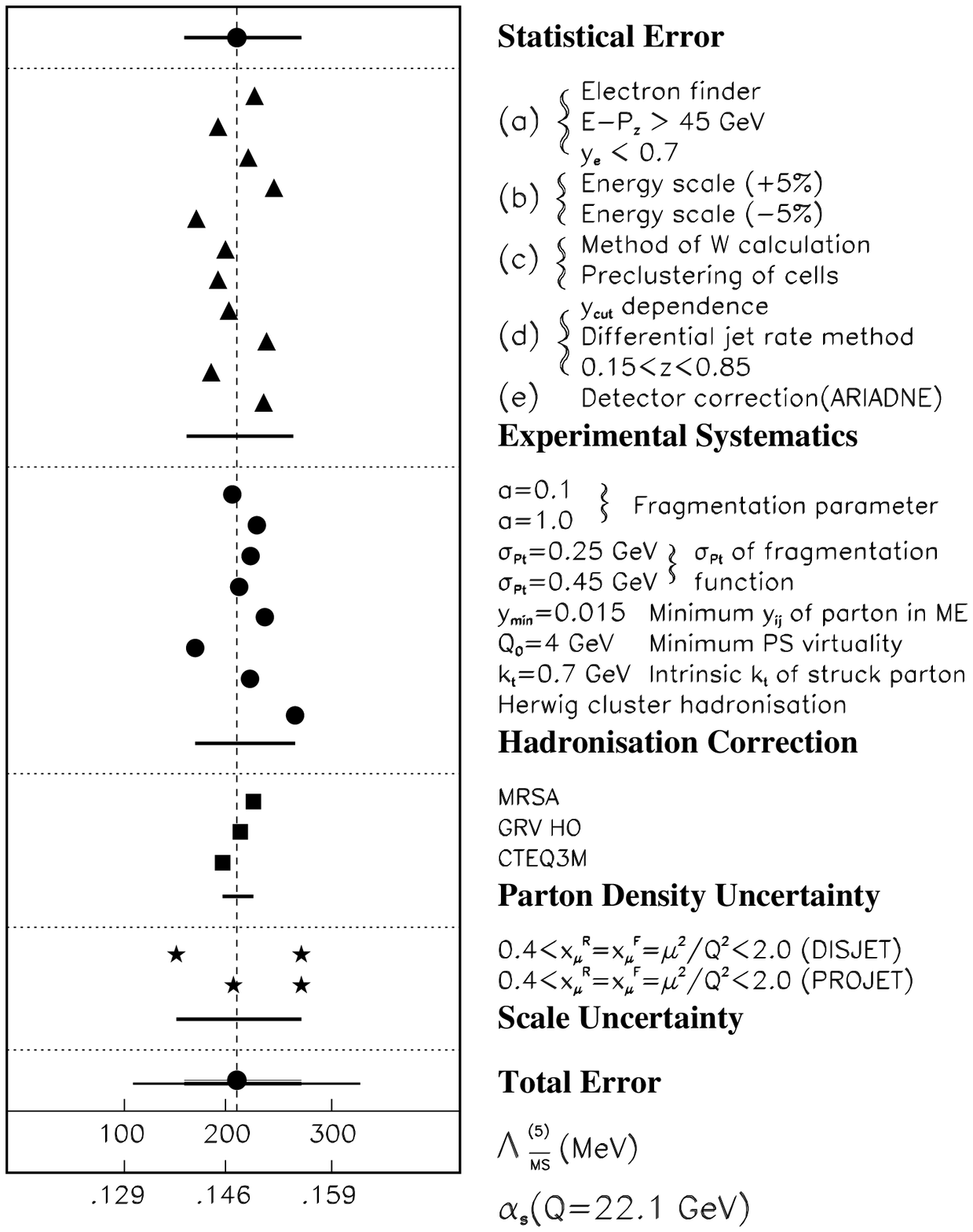,height=19.0cm,width=16.0cm}}
\caption{ Systematic uncertainties in the measured value of
\as\ (and $\lms^{(5)}$)
for $Q^2$ in the range $120<Q^2<3600$~GeV$^2$ expressed as the
deviation from the central value for the listed alterations in the analysis.
Sources of systematic uncertainties are grouped
into the four areas: experiment, hadronisation correction,
parton density, and scale. The experimental uncertainty is subdivided into:
(a) event selection, (b) energy scale, (c) jet analysis, (d) fitting
method, and (e) model dependence of the detector correction.
}
\label{fig:syst}
\end{figure}

Sources of systematic uncertainties in the \as\ determination were grouped
into the following classes: event selection, energy scale,
jet analysis, fitting
method, model dependence of detector corrections, hadronisation corrections
and parton density
(see Fig.~\ref{fig:syst}).
The first five classes were attributed to the
experimental systematic error. The uncertainties were studied for each of
the three $Q^2$ ranges separately as well as for the combined kinematic region.
Only the systematic uncertainty from the latter study is described here.
To illustrate the systematic uncertainty in \as\  associated with each
systematic effect, the fitted \as\ value obtained when the systematic
effect is varied was compared with the central value of
\as\ (see Tab.~\ref{tab:as}).

The subgroups of experimental
systematic errors are denoted by (a)-(e).
The systematic errors from the event selection (a) included: effect of using a
different electron finding algorithm; variations of the selection criteria,
$E-P_Z>45$ GeV; $y_e < 0.7$.
The errors from the energy scale (b) included a $\pm 5\%$ error assigned
to the uncertainty of the calorimeter energy response.
The errors from the jet analysis (c) included:
the choice of a different mass
scale, $W^2_{DA}= s\,(1-x_{DA})\,y_{DA}$  and
$W^2_{JB}=s\,(1-x_{JB})\,y_{JB}$,
in the JADE scaled mass definition, $y_{ij}=m^2_{ij}/W^2$;
the cells around the FCAL beam pipe, which contain mainly the proton remnant,
were first preclustered and the resulting objects were used in the jet
clustering algorithm (instead of the cell vectors themselves).
The errors from the fitting method (d) included: a QCD fit at
\ycut=0.03 instead of \ycut=0.02;
the analysis was cross-checked by a QCD fit to the differential
jet rates, $D_{1+1}$, defined by
$D_{1+1}(\ycut)=[R_{1+1}(\ycut+\Delta\ycut)-R_{1+1}(\ycut)]/\Delta\ycut$;
a more restrictive $z$-cut, $0.15<z<0.85$, was used.
Finally, the error from the model dependence
of the corrections for the detector acceptance and resolution (e)
was estimated by using the
colour-dipole model \cite{CDM} as implemented in the ARIADNE~4.06 Monte Carlo
\cite{ARIADNE4}.
The largest uncertainties for each subgroup were added in quadrature
to give the experimental systematic error.

To evaluate the uncertainty of the hadronisation correction, several
aspects of the hadronisation scheme were varied while the standard
detector corrections based on the LEPTO MEPS Monte Carlo were retained.
These studies were performed at the generator level.
First, parameters in the Lund
string fragmentation model \cite{JETSET} were varied:
$a$ in the `symmetric fragmentation
function', which regulates the longitudinal quark fragmentation,
was varied between 0.1 and 1; $\sigma_{Pt}$, which controls the hadron
transverse momentum distribution was varied between 0.25 and 0.45 GeV.
Second, parameters of the parton shower model
employed in the LEPTO MEPS Monte Carlo were changed: $y_{min}$
was varied between 0.005 and 0.015;
the minimum virtuality scale, $Q_0$, at which the parton showering is
stopped, was changed from 0.8 to 4~GeV; the primordial transverse
momentum $k_T$ of the struck parton in the proton was varied from 0.44 to
0.7~GeV.
Finally, a
completely different hadronisation model as
implemented in the  HERWIG~5.8 Monte
Carlo \cite{HERWIG} was used.  Most of these changes
result in relatively small systematic errors in  \as\
as shown in Fig.~\ref{fig:syst}.
The two largest deviations from the central value of \as\ arise from
the change of the hadronisation model and from the variation of $Q_0$.

We also repeated the analysis with parton density sets MRSA, GRV~HO, and
CTEQ~3M in the NLO calculation, all of which describe the results from
present DIS data well \cite{zeusf2}.
The differences in
\as (22.1~GeV)  from the central value are small ($<$~0.0022),
as shown in Fig.~\ref{fig:syst}. The fitted
\as\ value depends only weakly on the \as\ value used in the parton
density parametrisations \cite{VOGT}.

In the $x,Q^2$ region under study the contribution from massive $b-$quarks
to the proton structure function is calculated to be below 2\%  \cite{GRS}.
The effect of calculating with four instead of five flavours at the
BGF vertex was estimated with the PROJET program and was found to increase
$\as(\mz)$ by 0.0025. This number is not included in the systematic errors
given.

\subsection{Scale Dependence Effects}
\label{sec:discussion}

Our best estimate of the scale uncertainty in
the measured \as\ was obtained from DISJET and PROJET
by varying $\mu_R$ and $\mu_F$ from $0.4~Q^2$ to  $2.0~Q^2$,
redoing the fit to the jet rates and evolving
to obtain the corresponding value of \as\ at the original scale $Q$ (shown in
Fig.~\ref{fig:syst} for the full $Q^2$ range).
The scale dependence decreases with increasing $Q$ and
becomes negligible in the highest $Q^2$ interval. It is
slightly larger in DISJET than in PROJET.

Deep inelastic production of jets is a multi-scale process and it is not
evident that $Q^2$ is the best choice \cite{Brodkorb94,SCALES}
for the renormalisation and factorisation scales in the perturbative
calculation. Alternative scales have been suggested,
e.g. $p_T^2$ of the jets or the square of the invariant mass, $m_{ij}^2$,
of the two jets.
As a simple test the ratios $<p_T^2/Q^2>$ and $<m_{ij}^2/Q^2>$
were evaluated for the full $Q^2$ range for 2+1 jet events
and were used to estimate the resultant change of scale and hence of the
uncertainty in \as .
For our events these ratios typically lie between 0.4 and 2, i.e.
within the range explored above in our estimation of the scale uncertainty
using DISJET and PROJET.

For each group in Fig.~\ref{fig:syst} we quote the largest deviation from the
central value in each direction as the systematic error. The positive and
negative deviations are then added in quadrature separately
to give the systematic error.
The total systematic uncertainty for the value of \as\ resulting from
the effects
studied in Fig.~\ref{fig:syst} is comparable to the statistical errors.
In the final result given below the uncertainties from the experimental and
theoretical systematic effects (hadronisation, parton density distributions
and scale effects) are quoted separately.

\section{Summary}
\label{sec:conclusions}

Multi-jet production in $ep$ collisions was investigated using the
JADE jet
definition. In $ep$ collisions the application of the single jet
resolution parameter, \ycut ,  is
not sufficient to restrict the phase space of 2+1 jet production  to an
experimentally well understood and theoretically safe region. An
additional cut on the parton variable $z$ is introduced, which
excludes the problematic region where higher order effects are important
and jets are not well measured in the experiment.
With this additional cut the multi-jet production rate in DIS is
well reproduced by ${\cal O}(\as^2)$ perturbative QCD calculations.

The value of $\as(Q)$ was determined in three $Q^2$ regions in a single
experiment and was found to decrease with $Q$, consistent with the
running of the strong coupling constant.

The value for \as\ using the data from the entire kinematic range
$120<Q^2<3600$~GeV$^2$, and expressed at the $Z^0$ mass is given by:
\begin{eqnarray}
 \as (\mz) & = & 0.117~\pm~0.005~(stat)~^{+0.004}_{-0.005}~(exp)
                 ~^{+0.005}_{-0.004}~(had)~^{+0.001}_{-0.001}~(pd)
                 ~^{+0.005}_{-0.006}~(scale) \nonumber \\
   & = & 0.117~\pm~0.005~(stat)~^{+0.004}_{-0.005}~(syst_{exp})~
\pm 0.007 ~(syst_{theory}), \nonumber
\end{eqnarray}
 where {\it stat} corresponds to the statistical error and the systematic
 error components ({\it syst}) consist of the experimental ({\it exp}),
hadronisation ({\it had}),
parton density ({\it pd}) and the scale ({\it scale})
related uncertainties.
The overall systematic error is separated into its experimental
({\it exp}) and theoretical ({\it theory}) contribution.

Our value of \as\ is consistent with the most recent compilation
by the Particle Data Group \cite{PDG_as} of
previous measurements of $\as (\mz)$ using different methods:
$0.112 \pm 0.005$ (DIS),
$0.121 \pm 0.006$ ($e^+e^-$ event shape analysis)
and $0.124 \pm 0.007$ ($Z^0$ width).
The good agreement between our value of \as\ and
the results obtained using other methods in different
kinematic regimes represents a significant test of QCD.

\section*{Acknowledgements}
We thank the DESY directorate for their strong support and encouragement, and
the HERA machine group for their remarkable achievement in providing colliding
$ep$ beams.

It is a pleasure to thank E.~Mirkes and D.~Graudenz for providing NLO codes,
incorporating experimental needs, and for enlightening
discussions and comments. We  also wish to thank S.~Bethke, S.~Catani,
G.~Ingelman and A.~Vogt for fruitful discussions and comments.



\newpage

\end{document}